\documentclass[fleqn,usenatbib,onecolumn]{mnras}

\usepackage{newtxtext,newtxmath}

\usepackage[T1]{fontenc}
\usepackage{booktabs}
\DeclareRobustCommand{\VAN}[3]{#2}
\let\VANthebibliography\thebibliography
\def\thebibliography{\DeclareRobustCommand{\VAN}[3]{##3}\VANthebibliography}
\usepackage{amsfonts}
\newcommand{\msun}{M_{\odot}} 
\newcommand{\arepo}{\textsc{arepo}}

\usepackage{graphicx}	
\usepackage{amsmath}	

\providecommand{\apjl}{ApJ Lett.}
\providecommand{\apjs}{ApJS}
\providecommand{\aap}{A\&A}

\title[Gravitational lensing by a spiral galaxy]{Gravitational lensing by a spiral galaxy I: the influence from bar's structure to the flux ratio anomaly}

\author[X. Shan et al.]{
Xikai Shan,$^{1}$\thanks{E-mail: xk\_shan@mail.bnu.edu.cn}
Yunpeng Jin,$^{2}$
and Shude Mao$^{2}$
\\
$^{1}$Department of Astronomy, Tsinghua University, Beijing 100084, China\\
$^{2}$Department of Astronomy, Westlake University, Hangzhou 310030, Zhejiang Province, China\\
}

\date{Accepted XXX. Received YYY; in original form ZZZ}

\pubyear{\the\year{}}

\begin{document}
\label{firstpage}
\pagerange{\pageref{firstpage}--\pageref{lastpage}}
\maketitle

\begin{abstract}
Gravitational lens flux ratio anomalies are a powerful probe of small-scale mass structures, often attributed to dark matter subhalos. However, baryonic components can also play a significant role. This study investigates, for the first time, the impact of bars on flux ratio anomalies. We conduct a systematic analysis using barred galaxies from the Auriga simulations. First, we model the projected mass distribution with the Multi-Gaussian Expansion formalism. This method yields smooth lens potentials that preserve the primary bar structure while mitigating numerical noise. We then perform strong lensing simulations and quantify flux ratio anomalies by measuring their deviation from the theoretical cusp-caustic relation, denoted as $R_{\text{cusp}}$. 
Our primary finding is a strong, statistically significant correlation between the flux ratio anomaly magnitude and the strength of higher-order even Fourier modes. Specifically, the strengths of the boxy/peanut and hexapole components show an exceptionally tight correlation with $R_{\text{cusp}}$, with Spearman correlation coefficients of $r = 0.85$ and $0.89$, and p-values on the order of $10^{-6}$ and $10^{-8}$, respectively. This demonstrates that flux ratio anomalies are highly sensitive to complex, non-axisymmetric bar features. We conclude that flux ratio anomalies can be powerful indicators of bar morphology. Failing to account for such morphology can lead to misinterpreting lensing signatures and potentially overestimating the dark matter subhalo population.
\end{abstract}

\begin{keywords}
gravitational lensing: strong -- galaxies: bar
\end{keywords}


\section{Introduction} 
\label{sec:intro}
Gravitational lensing is a powerful tool for probing the distribution of matter, from individual galaxies to massive clusters~\citep{1992grle.book.....S, 1992ARA&A..30..311B}. In the strong lensing regime, multiple images of a background source are formed. The flux ratios between these images provide a direct test of the smoothness of the foreground lens's gravitational potential~\citep{1998MNRAS.295..587M, 2004ApJ...604L...5M}. When the observed fluxes deviate from the predictions of a simple, smooth lens model, this phenomenon is known as a ``flux ratio anomaly''. Such anomalies point to the existence of additional mass structures that perturb the lens potential on small scales.

These anomalies have several potential origins. First, they can be caused by perturbations from dark matter subhalos within the main lens galaxy. This idea was first proposed by~\cite{1998MNRAS.295..587M} and has been studied extensively with theoretical models~\citep{2001ApJ...563....9M, 2002ApJ...572...25D} and high-resolution N-body cosmological simulations~\citep{2009MNRAS.398.1235X, 2010MNRAS.408.1721X, Xu_2015}. Second, microlensing by the dense field of stars in the lens galaxy can contribute to the anomalies. This effect is particularly significant for lensed quasars observed at optical wavelengths~\citep{1989AJ.....98.1989I}. Third, other structures along the line of sight, such as the external shear from the large-scale environment or intervening halos, can alter the observed fluxes~\citep{2010ApJ...711..201S, 2012MNRAS.426.2978I, 2012MNRAS.421.2553X}. A fourth potential origin is the lens galaxy's own baryonic structures, such as edge on disks~\citep{2016MNRAS.463L..51H, 2017MNRAS.469.3713H, 2018MNRAS.475.2438H}.
Finally, multipole moments in strong lens galaxies can also contribute to flux-ratio anomalies~\citep{2024MNRAS.531.3431C}.

In this study, we focus on the impact of baryonic structures on flux ratio anomalies. Specifically, we investigate the effect of bars within spiral lens galaxies. Lensing by spiral galaxies offers several advantages over the more commonly studied elliptical lens systems. First, they enable a powerful combination of lensing constraints with kinematic data from the galaxy’s stellar or gas components, often obtained with Integral Field Unit (IFU) spectroscopic surveys. This synergy is crucial for breaking inherent degeneracies in dynamical studies, such as the disk-halo degeneracy~\citep{2000ApJ...533..194M, 2011MNRAS.417.1621D, 2012ApJ...750...10S}. Second, the multiply-imaged configurations from these lenses serve as a high-resolution diagnostic tool, allowing us to probe the complex internal structures of spiral galaxies, including their spiral arms and central bars.

Although the predicted strong lensing rate by spiral galaxies was once considered low~\citep{1998ApJ...495..157K}, a growing sample of these valuable systems is being discovered, thanks to ongoing and future large-scale surveys~\citep{2025NatAs...9.1116L}, such as Sloan WFC edge on Latetype Lens Survey (SWELLS)~\citep{2011MNRAS.417.1601T}, the Ultraviolet Near Infrared Optical Northern Survey (UNIONS)~\citep{2025arXiv250310610A}, the Euclid space telescope~\citep{2025arXiv250315324E}, the Vera C. Rubin Observatory~\citep{2009arXiv0912.0201L}, and the Chinese Space Station Telescope (CSST)~\citep{2024MNRAS.533.1960C, 2025arXiv250704618C}. These discoveries are further enhanced by high-resolution observations with the James Webb Space Telescope (JWST), which can precisely characterize the lens and source properties~\citep{2022A&A...666L...9C, 2025arXiv250308777N}. This influx of data ensures that the sample of spiral galaxy lenses will continue to grow significantly.
Based on the Euclid Q1 detection catalog, approximately 500 high-quality galaxy–galaxy strong lenses have been identified, of which 40 are classified as edge-on disk lenses. This corresponds to an observed edge-on disk fraction of approximately $8\%$ (40/500). By extrapolating this ratio to the total predicted Euclid yield of 110,000–120,000 lenses~\citep{2025arXiv250315324E,2025NatAs...9.1116L}, Euclid is expected to detect approximately 9,000 edge-on disk lenses.

This paper is the first in a series dedicated to exploring the influence of a key, yet under-investigated, structural component on flux ratio anomalies: the bar in galaxy. Photometric studies from optical to near-infrared wavelengths show that bars exist in more than half of nearby disk galaxies~\citep{2000AJ....119..536E, 2000ApJ...529...93K, 2007ApJ...659.1176M, 2007ApJ...657..790M, 2008ApJ...675.1194B, 2008ApJ...675.1141S, 2009A&A...495..491A, 2011MNRAS.411.2026M, 2015ApJS..217...32B, 2018MNRAS.474.5372E, 2025arXiv250502917J}. As a significant non-axisymmetric mass component in the galaxy's center, the bar is a natural candidate for perturbing the lens potential. This study, therefore, addresses two fundamental questions: Can bars produce measurable flux ratio anomalies? And conversely, can we use flux ratio anomaly measurements to infer the properties of bars in distant lensed galaxies? While previous work has explored the influence of angular multipoles on lensing observables~\citep{2003MNRAS.345.1351E, 2005MNRAS.364.1459C, 2024MNRAS.528.1757O, 2025PhRvD.111l3014P}, these studies have typically focused on the global structure of elliptical galaxies. Our work offers a new perspective by focusing specifically on the impact of bars within spiral lenses.

This paper is structured as follows. In Section~\ref{sec:data_pre}, we introduce the suite of galaxy simulations used in our analysis. Section~\ref{sec:method} details our lensing simulation and analysis methodology. We present our primary results in Section~\ref{sec:result}, followed by a summary and discussion of their implications in Section~\ref{sec:conclu_discus}.

\section{Data preparation} 
\label{sec:data_pre}

In this section, we describe the spiral galaxy data used in our strong lensing simulations (Section~\ref{subsec:auriga}) and detail the method used to mitigate shot noise (Section~\ref{subsec:mge_fit}).

\subsection{Auriga data projection}
\label{subsec:auriga}
The Auriga simulations~\citep{2017MNRAS.467..179G, 2024MNRAS.532.1814G} are a suite of high-resolution, cosmological zoom-in simulations of thirty Milky Way-mass galaxies ($M_{200} \approx 1-2 \times 10^{12} \msun$). These simulations were performed with the moving-mesh code \arepo{}, achieving a baryonic mass resolution of $\sim 5 \times 10^4 \msun$ and a dark matter resolution of $\sim 3 \times 10^5 \msun$. The adaptive gravitational softening length reaches a minimum of $\sim 370~\mathrm{pc}$. The simulations incorporate a comprehensive galaxy formation model that accounts for star formation, supernova (Type Ia/II) feedback, chemical evolution, active galactic nucleus (AGN) feedback, magnetic fields, radiative cooling, and a uniform UV background. This model was calibrated to match key observables, such as the stellar mass-halo mass relation.

For this study, we use a subset of 21 barred galaxies from the full sample of thirty, following the classification in \cite{2020MNRAS.491.1800B}. We use the $z=0$ snapshots and place the lens galaxies at a redshift of $z_l = 0.5$ and the background source at $z_s = 1.0$.
We also test whether the results depend on the choice of snapshot (i.e., the stage of galaxy evolution), and find that this does not affect our conclusions.
See Appendix~\ref{sec:ap_evo}.

The first step in our lensing simulation is to create a two-dimensional projected mass density map for each lens galaxy. We achieve this by mapping the simulation particles onto a 2D grid using a standard Smoothed Particle Hydrodynamics (SPH) kernel~\citep{1992ARA&A..30..543M}. During this step, we apply an adaptive smoothing technique to reduce shot noise from the discrete particles while preserving the primary bar structure. The smoothing length for each particle is set by the distance to its 640 nearest neighbors, with a maximum value capped at 1~kpc (the minimum softening length is $\sim$370 pc). While this initial smoothing affects the projected structure, it does not bias our final analysis. We have conducted a test using fewer nearest neighbors and found that the final conclusion is not influenced, though the SPH density projection map, as shown in Figure~\ref{fig:halo_10}, will exhibit more numerical noise.
The reason our conclusion is not influenced is that we use a self-consistent method to describe the bar's structure and the bar-induced flux ratio anomaly, so our method for quantifying flux ratio anomalies inherently accounts for the smoothing scale.
One can see the text below for more details.

Figure~\ref{fig:halo_10} shows the projection of a representative Auriga galaxy (Au 10). The x and y axes are the spatial coordinates in kiloparsecs (kpc). The left and right columns display the galaxy from face on and edge on viewing angles, respectively. The color scale indicates the dimensionless convergence, $\kappa$, which is the two-dimensional mass density scaled by the critical surface density, $\Sigma_\mathrm{crit} = \frac{c^{2}}{4 \pi G} \frac{D_{\mathrm{s}}}{D_{\mathrm{l}} D_{\mathrm{ls}}}$. The white lines are isodensity contours.

As seen in the figure, a bar structure is clearly visible in the galaxy's center, especially in the side-on projection. However, the isodensity contours are also irregular and asymmetric. These features arise from a combination of residual shot noise and physical substructures within the galaxy. To isolate the influence of the bar on flux ratio anomalies, we must mitigate these small-scale fluctuations. This ensures that any measured anomaly is driven primarily by the bar's structure, not by other factors. Therefore, we use the Multi-Gaussian Expansion (MGE) fitting method to create a smoothed representation of the galaxy's surface brightness/surface density.

\subsection{Multi-Gaussian fitting}
\label{subsec:mge_fit}
To address the presence of shot noise and small-scale fluctuations in the simulated convergence maps, we utilize an MGE fitting technique~\citep{1994A&A...285..723E, 2002MNRAS.333..400C}. 
This method involves modeling the projected mass density, $\kappa(R', \theta')$, as a superposition of several concentric, two-dimensional Gaussian functions:
\begin{equation}
\label{eq:mge_fit}
\kappa \left(R^{\prime}, \theta^{\prime}\right)=\sum_{j=1}^{N} \frac{\kappa_{j}}{2 \pi \sigma_{j}^{\prime 2} q_{j}^{\prime}} \exp \left[-\frac{1}{2 \sigma_{j}^{\prime 2}}\left(x_{j}^{\prime 2}+\frac{y_{j}^{\prime 2}}{q_{j}^{\prime 2}}\right)\right]
\end{equation}
In the two-dimensional plane, $(R', \theta')$ are the polar coordinates. 
The coordinates $(x'_{j}, y'_{j})$ are Cartesian coordinates, given by $x_{j}^{\prime}=R^{\prime} \sin(\theta^{\prime}-\psi_{j})$ and $y_{j}^{\prime}=R^{\prime} \cos(\theta^{\prime}-\psi_{j})$.
The parameters for each Gaussian are:
\begin{itemize}
    \item $\kappa_{j}$: The central convergence of the $j$-th component.
    \item $\sigma'_{j}$: The dispersion along the major axis.
    \item $q'_{j}$: The axial ratio.
\end{itemize}

\begin{figure} 
\centering 
\includegraphics[width=\columnwidth]{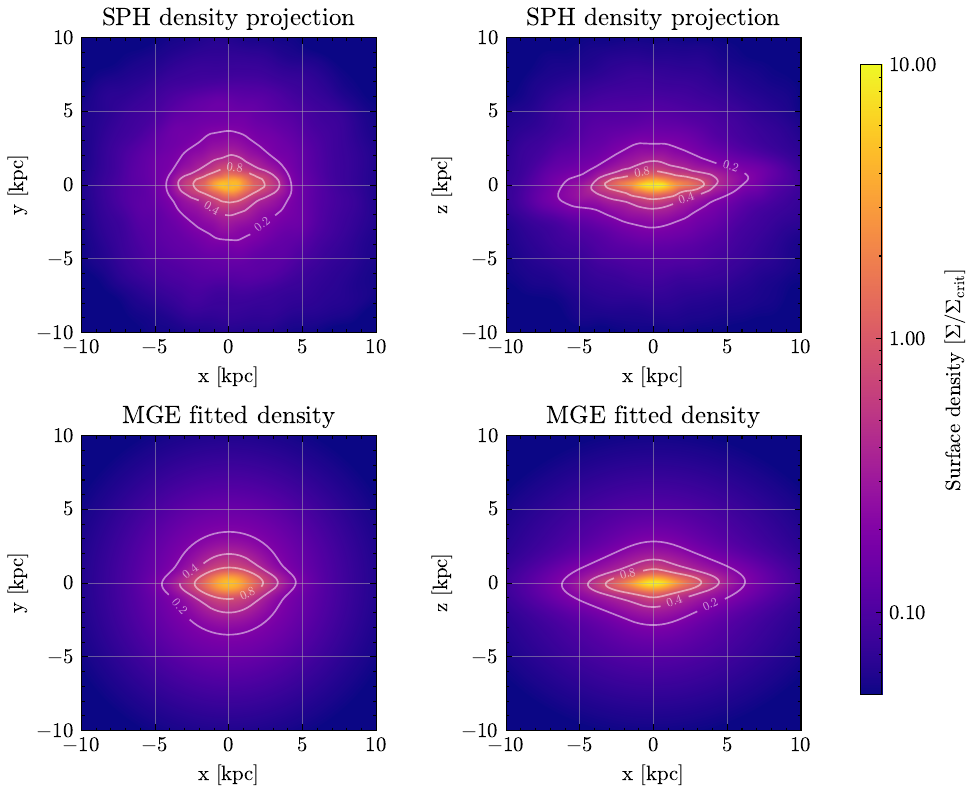} 
\caption{This figure displays the dimensionless surface density maps ($\kappa = \Sigma/\Sigma_\mathrm{crit}$) for Au 10 from the Auriga simulations. The columns show two different projections: a face on view (left) and an edge on view with the bar oriented side-on (right). The top panels present the density maps derived from the direct projection using an SPH kernel, while the bottom panels show the corresponding maps reconstructed using a Multi-Gaussian Expansion (MGE) fit. In all panels, the color bar indicates the density level, with warmer colors denoting higher density regions, and the white lines represent iso-density contours.
} 
\label{fig:halo_10} 
\end{figure}

The bottom panel of Figure~\ref{fig:halo_10} displays the MGE fitting results for the Auriga galaxy (Au 10).
Compared with the direct SPH projection in the upper panel, the isodensity contours (white curves) from the MGE fit are smoother and more symmetric.
More importantly, the central bar structure is clearly resolved.

Therefore, the MGE method effectively smooths the density distribution while preserving key bar structural features. 
Based on this result, we will construct the lensing system using only the two-dimensional density distribution obtained from the MGE fitting.

\section{Lensing methodology} 
\label{sec:method}

\subsection{Lensing basic theory}
\label{subsec:lens_the}

The lens equation for a single lens plane can be written as:
\begin{equation}
\label{eq:lens_eq}
\mathbf{y} = \mathbf{x} - \boldsymbol{\alpha}(\mathbf{x}),
\end{equation}
where $\mathbf{y}$ is the position of the source in the source plane, $\mathbf{x}$ is the corresponding image position in the lens plane, and $\boldsymbol{\alpha}(\mathbf{x})$ is the deflection angle. The critical quantity in this equation is the deflection angle, which is the gradient of the two-dimensional lensing potential $\psi(\mathbf{x})$:
\begin{equation}
\label{eq:alpha}
\boldsymbol{\alpha}(\mathbf{x}) = \boldsymbol{\nabla} \psi(\mathbf{x}).
\end{equation}

The potential is related to the convergence $\kappa(\mathbf{x})$ of the lens galaxy  through the Poisson equation:
\begin{equation}
\label{eq:poisson}
\nabla^{2} \psi(\mathbf{x}) = 2 \kappa(\mathbf{x}).
\end{equation}

As introduced in Section~\ref{sec:data_pre}, the convergence map has already been generated. The next critical step is to solve Eq.~(\ref{eq:poisson}) to derive the lensing potential. We achieve this using a standard Fast Fourier Transform (FFT) method. The convergence map has a resolution of 0.01 kpc, which is substantially smaller than the typical size of a bar (a few kpc), and a boundary of 20 kpc, corresponding to approximately 0.1 $R_{200}$. $R_{200}$ is the radius within which the mean density of the halo equals 200 times the cosmic critical density, $\rho_c(z)$, where z is the redshift of the halo.
This choice represents a balance between precision (resolution effect) and accuracy (boundary effect) under the constraints of current computational resources.
Although the truncation of the density map can introduce external shear, the shear amplitude is only on the order of $10^{-2}$ at this truncation radius. 
Therefore, we neglect this effect on the flux ratio anomaly measurement~\citep{2020A&A...644A.108V}.
Additionally, the finite resolution of the density map affects the accuracy of magnification calculations, particularly in regions of very high magnification. 
These regions occur in the proximity of critical curves, where the image magnification, $\mu$, is inversely proportional to its distance, $\delta r$, from the critical curve (i.e., $\mu \propto 1/\delta r$; \cite{1992grle.book.....S}).
Given the resolution of our density map, we therefore impose an artificial magnification threshold of $\mu=100$.


Once the lensing potential $\psi(\mathbf{x})$ is obtained, it can be substituted into Eq.~(\ref{eq:alpha}) and Eq.~(\ref{eq:lens_eq}) to solve for the image positions $\mathbf{x}$ for a specified source position $\mathbf{y}$. 
After locating all images, their corresponding magnifications, $\mu$, are calculated by using:
\begin{equation} 
\label{eq:magnification}
\mu = \frac{1}{(1-\kappa)^2 - \gamma^2},
\end{equation}
where $\kappa$ is the convergence and $\gamma$ is the shear magnitude. 
These quantities are defined in terms of the second partial derivatives of the lensing potential $\psi_{ij} = \frac{\partial^2\psi}{\partial x_i\partial x_j}$.
The convergence is given by:
\begin{equation} 
\label{eq:convergence}
\kappa = \frac{1}{2}(\psi_{11} + \psi_{22}),
\end{equation}
and the shear is characterized by two components, $\gamma_1$ and $\gamma_2$:
\begin{align} 
\label{eq:shear_components}
\gamma_1 &= \frac{1}{2}(\psi_{11} - \psi_{22}), \\
\gamma_2 &= \psi_{12} = \psi_{21}.
\end{align}
The shear magnitude is the combination of these two components $\gamma = \sqrt{\gamma_1^2 + \gamma_2^2}$.

\subsection{Cusp caustic relation}
\label{subsec:R_cusp}
In a strong gravitational lensing system, a source located near a caustic cusp produces three images on one side of the lens center. 
For a smooth lens potential, the magnifications of these three images are expected to satisfy the theoretical cusp caustic relation:
\begin{equation}
\label{eq:R_cusp}
\mathrm{R_{cusp}} \equiv \frac{|\mu_A + \mu_B + \mu_C|}{|\mu_A| + |\mu_B| + |\mu_C|} \to 0 \quad (\Delta \beta \to 0),
\end{equation}
where $\Delta \beta$ represents the distance from the source to the caustic cusp, and $\mu_A$, $\mu_B$, and $\mu_C$ are the magnifications of the three respective images.

\begin{figure} 
\centering 
\includegraphics[width=\columnwidth]{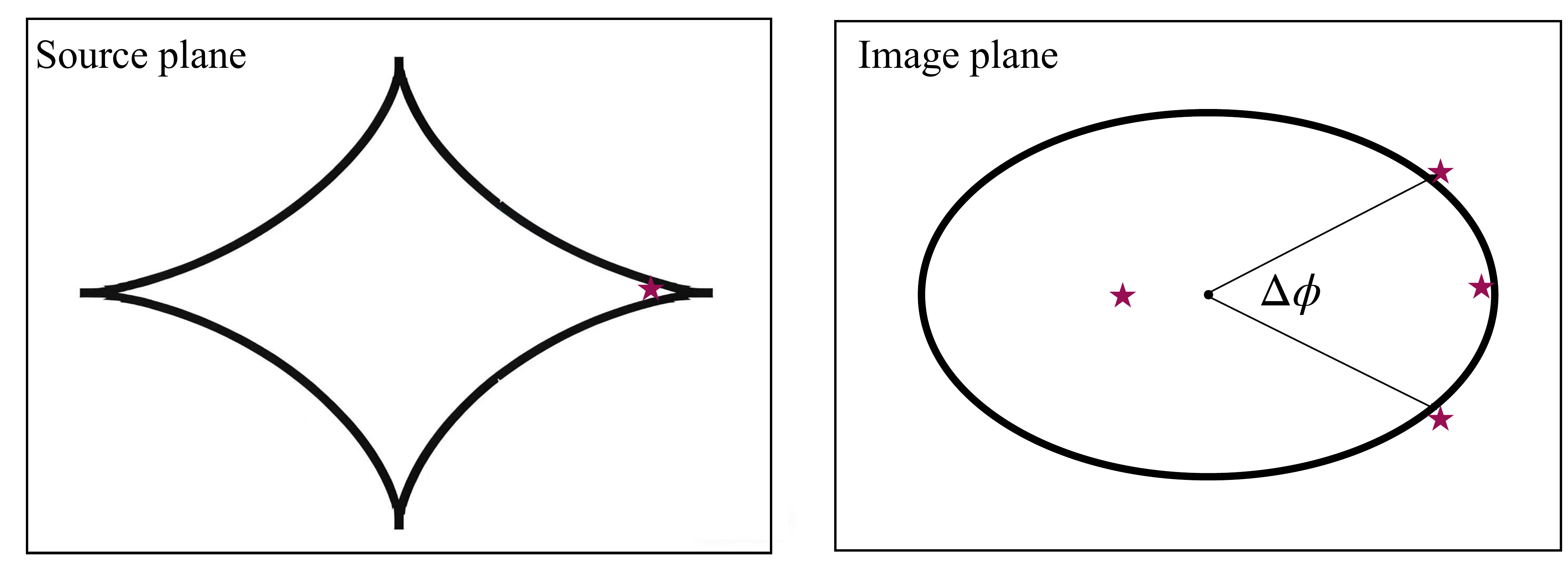} 
\caption{The left panel shows a tangential caustic (where the source image is stretched along the tangential direction) in the source plane, and the red star indicates a source located near the cusp region. The right panel shows the critical curve in the image plane, where one can see four strong lensing images corresponding to the source. Note that the size of the star does not represent the strong lensing magnification. In the right panel, the angle $\Delta \phi$ is the opening angle defined in Section~\ref{subsec:R_cusp}. }
\label{fig:skretch} 
\end{figure}

Since the source-caustic distance, $\Delta \beta$, is not directly observable in a strong lensing system, a proxy relationship between $\mathrm{R_{cusp}}$ and the opening angle of the images, $\Delta \phi$, is employed. 
The opening angle is defined as the angle formed by the two outermost images, with the vertex at the lens center. 
This angle can be measured directly from the lensed image, and a smaller opening angle indicates that the source is closer to the caustic cusp.
Consequently, the opening angle $\Delta \phi$ serves as a key observable for testing the cusp caustic relation.
For a more intuitive definition, please refer to Figure~\ref{fig:skretch}.

However, the assumption of a smooth lens potential is an idealization, as lens galaxies typically contain substructures. These substructures perturb the gravitational potential and can violate the cusp caustic relation, leading to a phenomenon known as a ``flux ratio anomaly''~\citep{1998MNRAS.295..587M}.
In this paper, we utilize this relationship to quantify the influence of a bar on the observed flux ratio anomalies.

\section{Result}
\label{sec:result}

In this section, we present the results of our lensing system construction and investigate the correlation between the strength of the angular complexity in the bar region and two key metrics: the caustic area, a proxy for lensing probability, and flux-ratio anomalies.

\subsection{Lens Mapping for Auriga Halos}
\label{subsec:lensing_mapping_auriga}

\begin{figure} 
\centering 
\includegraphics[width=\columnwidth]{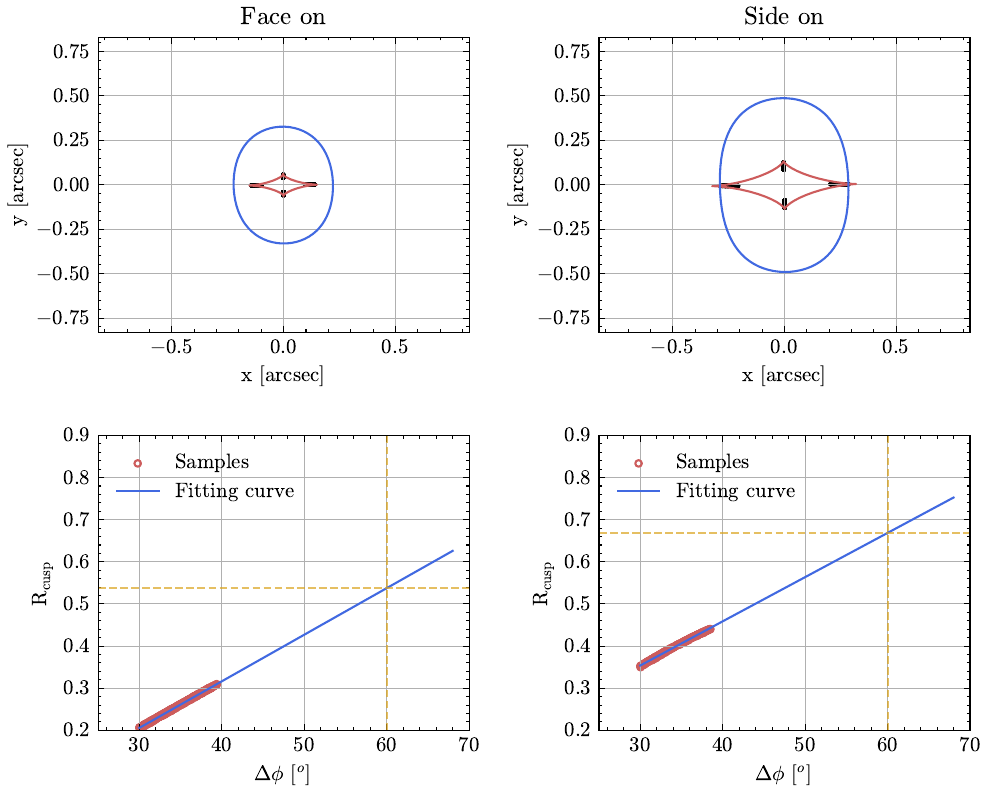} 
\caption{This figure displays the source sampling and $\mathrm{R_{cusp}}$ result for Au 10 from the Auriga simulations. The top panels illustrate the tangential (diamond-shaped, where the source image is stretched along the tangential direction) and radial (elliptical-shaped, where the source image is stretched along the radial direction) caustics.
The columns correspond to different galaxy projections: the left panel shows a face on view and the right panel shows an edge on view with a side on bar. 
The gray dots near the caustic cusps represent the sampled source positions, with $10^4$ points used for each position. 
The lower panels plot the cusp caustic relation, $\mathrm{R_{cusp}}$ (defined in Eq.~\ref{eq:R_cusp}), as a function of the opening angle, $\Delta \phi$. 
In these plots, the red open circles denote the data from the sampled points, while the solid blue curves represent the best-fit linear polynomials. 
The yellow cross in each lower panel indicates the value of $\mathrm{R_{cusp}}$ at an opening angle of $60^\circ$, which is the value we used to characterize the lensing flux ratio anomaly.
} 
\label{fig:halo_10_Cusp_caustic_relation} 
\end{figure}

Here, we illustrate the construction of the lensing system based on the Auriga simulated data. Figure~\ref{fig:halo_10_Cusp_caustic_relation} displays the results for Au 10, which is the same halo shown in Figure~\ref{fig:halo_10}. We set the lens redshift to $z_l=0.5$ and the source redshift to $z_s=1.0$.

The top panels of Figure~\ref{fig:halo_10_Cusp_caustic_relation} show the caustics for different projection angles: face on (left), edge on with the side-on bar (right). The x and y axes represent the source plane coordinates in units of arcseconds. The red curves denote the tangential caustics, which produce four images, while the blue curves are the radial caustics (also known as cuts), which produce two images. The gray shaded regions near the caustic cusps indicate the source sampling areas, where we randomly placed $10^4$ sources at each cusp.

The bottom panels show the corresponding $\mathrm{R_{cusp}}$ values, as defined in Eq.~(\ref{eq:R_cusp}), as a function of the opening angle, $\Delta \phi$, of the image triplets, with the vertex at the lens position. The red open circles represent the sample points. To ensure that the sources are located as close to the cusp as possible, we only include results with an opening angle of less than $90^\circ$.
Through this analysis, we find that sources near the minor cusps (the regions along the x-axis) readily produce triplet images with opening angles of less than $90^\circ$. Conversely, for positions near the major cusps (the regions along the y-axis), only a few points generate triplet images with opening angles smaller than $90^\circ$. This is because the major cusp is influenced by the underlying smooth elliptical disk or bar structure. At the resolution of our lensing convergence map, we are unable to simulate a source sufficiently close to the major cusp to generate images with such a small opening angle. Therefore, in the bottom panels and the subsequent analysis, we only consider the results from the minor cusp regions.

From the lower panels, it is evident that $\mathrm{R_{cusp}}$ is approximately linearly proportional to the opening angle $\Delta \phi$. 
Based on this property, we fit the data points with a linear curve. 
As shown, the blue fitted curve accurately represents the data. 
To quantify the effect of the bar across different scenarios consistently, we use the value of $\mathrm{R_{cusp}}$ at an opening angle of $60^\circ$ to represent the flux ratio anomaly. 
This value can be obtained from the yellow cross curves.


\begin{figure} 
\centering 
\includegraphics[width=0.7\columnwidth]{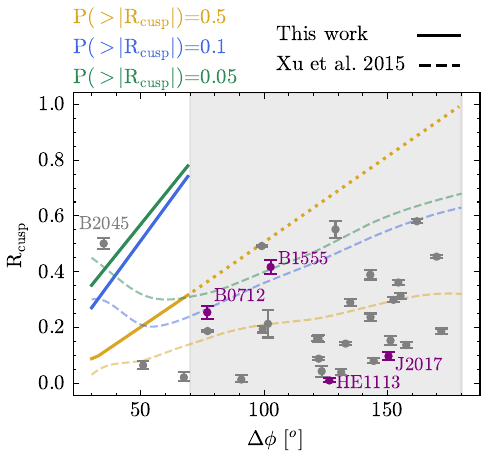} 
\caption{Contour plots of the conditional probability $P (>|\mathrm{R_{cusp}}|)$ for a given opening angle $\Delta \phi$.
Different colors represent different probability values: yellow (50\%), blue (10\%), and green (5\%).
The solid curves represent the results from the 21 barred galaxies from the Auriga simulations used in this work.
The dotted curve represents the predicted probability contour for $P (>|\mathrm{R_{cusp}}|)=0.5$, obtained using a linear fitting formula.
The dashed curves following the same color code represent the probability contours from Figure 4 of~\citet{Xu_2015}, which show results for smooth elliptical lens potentials with a subhalo population hosted by a group-sized halo ($M_{200} = 5\times10^{13}h^{-1}M_\odot$).
The circles with error bars represent the currently observed strongly lensed quasars with flux ratio measurements, as summarized in Table~\ref{tab:obs_rcusp} (31 in total).
This sample combines those presented in~\citet{Xu_2015, 2024MNRAS.530.2960N, Keeley2024_JWST_II}, and \citet{Keeley2025_JWST_III}.
The grey points denote elliptical lens galaxies, and the purple points denote spiral lens galaxies.}
\label{fig:R_cusp_JWST_sample} 
\end{figure}

Then, in Figure~\ref{fig:R_cusp_JWST_sample}, we show the contour maps of conditional probabilities $P (>|\mathrm{R_{cusp}}|)$ for a given opening angle $\Delta \phi$, by combining all 21 barred galaxies from the Auriga simulations. 
The solid curves in different colors represent different probability values: yellow (50\%), blue (10\%), and green (5\%). 
Here, we only show the results for opening angles less than 70$^\circ$. For larger angles, there are too few simulated cases to generate reliable results due to the sample limits of our simulations. 
However, we extend the $P (>|\mathrm{R_{cusp}}|)=50\%$ curve to $\Delta \phi=180^\circ$ using a linear fitting formula. This choice is motivated by the almost linear behavior seen in Figure 4 of \citet{Xu_2015}. 

In this figure, the dashed curves (using the same color code) show the results for smooth elliptical lens potentials with a subhalo population hosted by a group-sized halo ($M_{200} = 5\times10^{13}h^{-1}M_\odot$), taken from Figure 4 of \citet{Xu_2015}. 
We can see that the predicted $\mathrm{R_{cusp}}$ values in barred galaxies are generally larger than those in elliptical galaxies, even when subhalo effects are considered. 

Finally, we plot the currently observed strongly lensed quasars. 
We combine the samples presented in~\citet{Xu_2015, 2024MNRAS.530.2960N, Keeley2024_JWST_II} and \citet{Keeley2025_JWST_III}. 
Specifically,~\citet{Xu_2015} provides 8 radio lens samples. The recent JWST surveys \citep{2024MNRAS.530.2960N, Keeley2024_JWST_II, Keeley2025_JWST_III} observed 31 lenses in total using the mid-infrared band. This band specifically targets the warm-dust emission of quasars because its physical size (1-10 pc) is large enough to avoid stellar microlensing, but small enough to remain sensitive to dark matter subhaloes. After removing systems without clear quadruple images or reliable lens mass centers, we obtain 23 new viable samples. 
We summarize all these observed samples in Table~\ref{tab:obs_rcusp}.

In this Figure, the grey points represent elliptical lens galaxy samples, while the purple points represent spiral lens galaxies. 
We find that the $\mathrm{R_{cusp}}$ values of the four observed spiral lenses all fall below our 50\% probability contour. This indicates that their $\mathrm{R_{cusp}}$ values are relatively small and do not show strong anomalies; they can be well explained by the smooth macro-models of barred galaxies. 
In contrast, previous studies have pointed out that the dark matter subhalo populations predicted by standard cosmological simulations are generally not large enough to explain the extreme flux-ratio anomalies seen in some elliptical lenses~\citep{2009MNRAS.398.1235X, 2010MNRAS.408.1721X, Xu_2015}. 
Therefore, our result highlights that flux-ratio anomalies should be carefully examined using more realistic strong lensing macro-models (e.g., including galactic bars) before using them to constrain dark matter substructures in spiral lens systems.

\begin{table*}
	\centering
	\caption{Observed opening angles ($\Delta\phi$) and $R_{\mathrm{cusp}}$ values for the sample of quadruply lensed quasars used in this work. To avoid contamination from stellar microlensing, the $R_{\mathrm{cusp}}$ values and their associated uncertainties are derived exclusively from observations at radio or mid-infrared (warm-dust) wavelengths. Data are compiled from \citet{Xu_2015, 2024MNRAS.530.2960N, Keeley2024_JWST_II, Keeley2025_JWST_III} and references therein.}
	\label{tab:obs_rcusp}
	\begin{tabular}{l c c l} 
		\toprule
        
		Lens Name & $\Delta\phi$ [$^{\circ}$] & $\mathrm{R_{cusp}}$ & References \\
		\midrule
		B0128+437   & $123.3$ & $0.043 \pm 0.020$ & \citet{2003ApJ...595..712K, 2000MNRAS.319L...7P} \\
		MG0414+0534 & $101.5$ & $0.213 \pm 0.049$ & \citet{1999ApJ...523..617F, 1995AJ....110.2570L, 1997ApJ...475..512K, 2000AA...362..845R} \\
		B0712+472   & $76.9$  & $0.254 \pm 0.024$ & \citet{2003ApJ...595..712K, 1998MNRAS.296..483J, 2000MNRAS.311..389J} \\
		B1422+231   & $77.0$  & $0.187 \pm 0.004$ & \citet{2003ApJ...595..712K, 1996ApJ...462L..53I, 1999MNRAS.307L...1P, 1999ApJ...523..617F} \\
		B1555+375   & $102.6$ & $0.417 \pm 0.026$ & \citet{2003ApJ...595..712K, 1999AJ....118..654M} \\
		B1608+656   & $99.0$  & $0.492 \pm 0.002$ & \citet{1999ApJ...527..513K, 1996ApJ...460L.103F} \\
		B1933+503   & $143.0$ & $0.389 \pm 0.017$ & \citet{2001ApJ...554.1216C, 1998MNRAS.301..310S, 2000MNRAS.318...73B} \\
		B2045+265   & $34.9$  & $0.501 \pm 0.020$ & \citet{2003ApJ...595..712K, 1999AJ....117..658F, 2007MNRAS.378..109M} \\

		J0248       & $121.2$ & $0.063 \pm 0.012$ & \citet{2019MNRAS.483.5649S, 2019AA...622A.165D, Keeley2024_JWST_II, Keeley2025_JWST_III} \\
		J0259       & $70.8$  & $0.147 \pm 0.015$ & \citet{2018RNAAS...2...21S, Keeley2024_JWST_II, Keeley2025_JWST_III} \\
		J0405       & $108.9$ & $0.046 \pm 0.011$ & \citet{2018MNRAS.480.5017A, Keeley2024_JWST_II, Keeley2025_JWST_III} \\
		MG0414      & $100.8$ & $0.111 \pm 0.014$ & \citet{1992AJ....104..968H, Keeley2024_JWST_II, Keeley2025_JWST_III} \\
		HE0435      & $113.8$ & $0.133 \pm 0.009$ & \citet{2002AA...395...17W, Keeley2024_JWST_II, Keeley2025_JWST_III} \\
		J0607       & $128.5$ & $0.135 \pm 0.038$ & \citet{2021ApJ...921...42S, 2023MNRAS.520.3305L, Keeley2024_JWST_II, Keeley2025_JWST_III} \\
		J0608       & $113.1$ & $0.137 \pm 0.007$ & \citet{2021ApJ...921...42S, 2023MNRAS.520.3305L, Keeley2024_JWST_II, Keeley2025_JWST_III} \\
		J0659       & $96.8$  & $0.207 \pm 0.011$ & \citet{2019AA...622A.165D, 2023MNRAS.520.3305L, Keeley2024_JWST_II, Keeley2025_JWST_III} \\
		J0803       & $103.3$ & $0.046 \pm 0.008$ & \citet{2023MNRAS.520.3305L, Keeley2024_JWST_II, Keeley2025_JWST_III} \\
		J0924       & $78.0$  & $0.320 \pm 0.008$ & \citet{2003AJ....126..666I, Keeley2024_JWST_II, Keeley2025_JWST_III} \\
		HE1113      & $74.2$  & $0.008 \pm 0.008$ & \citet{2008AJ....135..374B, Keeley2024_JWST_II, Keeley2025_JWST_III} \\
		PG1115      & $86.5$  & $0.075 \pm 0.014$ & \citet{1980Natur.285..641W, Keeley2024_JWST_II, Keeley2025_JWST_III} \\
		RXJ1131     & $80.2$  & $0.086 \pm 0.016$ & \citet{2003AA...406L..43S, Keeley2024_JWST_II, Keeley2025_JWST_III} \\
		GRAL1131    & $108.3$ & $0.160 \pm 0.015$ & \citet{2018AA...616L..11K, Keeley2024_JWST_II, Keeley2025_JWST_III} \\
		2M1134      & $152.0$ & $0.155 \pm 0.004$ & \citet{2018MNRAS.476..927L, Keeley2024_JWST_II, Keeley2025_JWST_III} \\
		J1251       & $96.8$  & $0.044 \pm 0.016$ & \citet{2007AJ....134.1515K, Keeley2024_JWST_II, Keeley2025_JWST_III} \\
		H1413       & $94.6$  & $0.108 \pm 0.006$ & \citet{1988Natur.334..325M, Keeley2024_JWST_II, Keeley2025_JWST_III} \\
		J1537       & $89.0$  & $0.207 \pm 0.011$ & \citet{2018MNRAS.479.5060L, 2019AA...622A.165D, Keeley2024_JWST_II, Keeley2025_JWST_III} \\
		PSJ1606     & $97.1$  & $0.141 \pm 0.007$ & \citet{2018MNRAS.479.5060L, Keeley2024_JWST_II, Keeley2025_JWST_III} \\
		J2017       & $89.2$  & $0.138 \pm 0.013$ & \citet{2021ApJ...921...42S, Keeley2024_JWST_II, Keeley2025_JWST_III} \\
		WFI2033     & $90.9$  & $0.054 \pm 0.005$ & \citet{2004AJ....127.2617M, Keeley2024_JWST_II, Keeley2025_JWST_III} \\
		J2038       & $68.4$  & $0.065 \pm 0.012$ & \citet{2018MNRAS.479.4345A, Keeley2024_JWST_II, Keeley2025_JWST_III} \\
		J2205       & $113.8$ & $0.088 \pm 0.007$ & \citet{2023MNRAS.520.3305L, Keeley2024_JWST_II, Keeley2025_JWST_III} \\
		J2344       & $98.1$  & $0.076 \pm 0.011$ & \citet{2017AJ....153..219S, Keeley2024_JWST_II, Keeley2025_JWST_III} \\
		\bottomrule
	\end{tabular}
    
\end{table*}

\subsection{The description of the angular complexity strength in the bar region}
\label{subsec:bar_str}
In this section, we use a quantitative framework to characterize the structural properties of bars. Our primary objective is to develop the necessary tools to test the proposed connection between bar strength and flux ratios anomaly.

While the concept of a ``strong'' or ``weak'' bar is intuitive, its quantification is not trivial. A variety of metrics have been proposed in the literature, ranging from the bar's projected axial ratio to functions of the tangential forces exerted on galactic material \citep{1981A&A....96..164C, 2001A&A...375..761B}. In this work, we adopt the Fourier decomposition method, a powerful technique for isolating non-axisymmetric structures, as detailed in~\cite{2002MNRAS.330...35A}. This approach allows us to measure the strength of the bar and its associated higher-order components, such as boxy and peanut-shaped morphologies.

The method begins by decomposing the galaxy's deprojected surface density, $\Sigma(r, \theta)$, into a series of Fourier modes within concentric radial annuli:
\begin{equation}
    A_m(r) = \frac{1}{\pi} \int_0^{2\pi} \Sigma(r, \theta) e^{-i m\theta} \, d\theta, \quad m = 0, 1, 2, \ldots
    \label{eq:fourier_coeff}
\end{equation}
Here, $A_m(r)$ is the complex Fourier amplitude of the $m$-th component at radius $r$. The $m=0$ mode, $A_0(r)$, corresponds to the azimuthally averaged surface density at that radius.
The physical interpretation of other modes are also well-established:
The even-numbered modes trace the primary bisymmetric structures characteristic of barred galaxies.
\begin{itemize}
    \item \textbf{The $m=2$ mode:} This is the dominant mode in a barred galaxy and directly measures the primary strength of the bar itself~\citep{2002MNRAS.330...35A}.
    \item \textbf{The $m=4$ mode:} This mode quantifies the bar's deviation from a pure elliptical shape. A significant $m=4$ component is the classic signature of the boxy or peanut-shaped bulges that are often the result of vertical buckling instabilities in the bar~\citep{1981A&A....96..164C, 2005AIPC..804..333A}.
    \item \textbf{The $m=6$ mode:} A non-zero $m=6$ component can trace hexapole (six-fold) distortions in the central regions or the presence of an inner ring. It can also be associated with the bases of spiral arms emerging from the ends of the bar~\citep{2002MNRAS.330...35A}.
\end{itemize}

The odd-numbered modes describe asymmetric features within the galaxy. While generally weaker than the even modes in mature, isolated barred galaxies, they can provide important physical insights.
\begin{itemize}
    \item \textbf{The $m=1$ mode:} This mode measures the galaxy's ``lopsidedness'', corresponding to a large-scale asymmetry where the nucleus is displaced relative to the outer disk isophotes. Such features are often attributed to tidal interactions with companion galaxies or asymmetries in the accretion of gas or dark matter~\citep{1995ApJ...447...82R, 1997ApJ...477..118Z}.
    \item \textbf{The $m=3$ and $m=5$ modes:} These higher-order odd modes trace more complex asymmetries, such as triangular distortions ($m=3$)~\citep{1984PhR...114..319A}.
\end{itemize}

To quantify the relative contribution of each high order component ($m=1,2,3,\ldots$), we normalize its amplitude by the background $A_0(r)$. 
We then define the maximum value of the ratio $A_m(r) / A_0(r)$ across all radii within the bar region as a global measure of the strength of the $m$-th order structural component.

Finally, it is worth noting that to ensure the decomposition results reflect the bar's influence, we limit the calculation to the bar's region, as defined in Table 1 of~\cite{2020MNRAS.491.1800B}.

\begin{figure} 
\centering 
\includegraphics[width=\columnwidth]{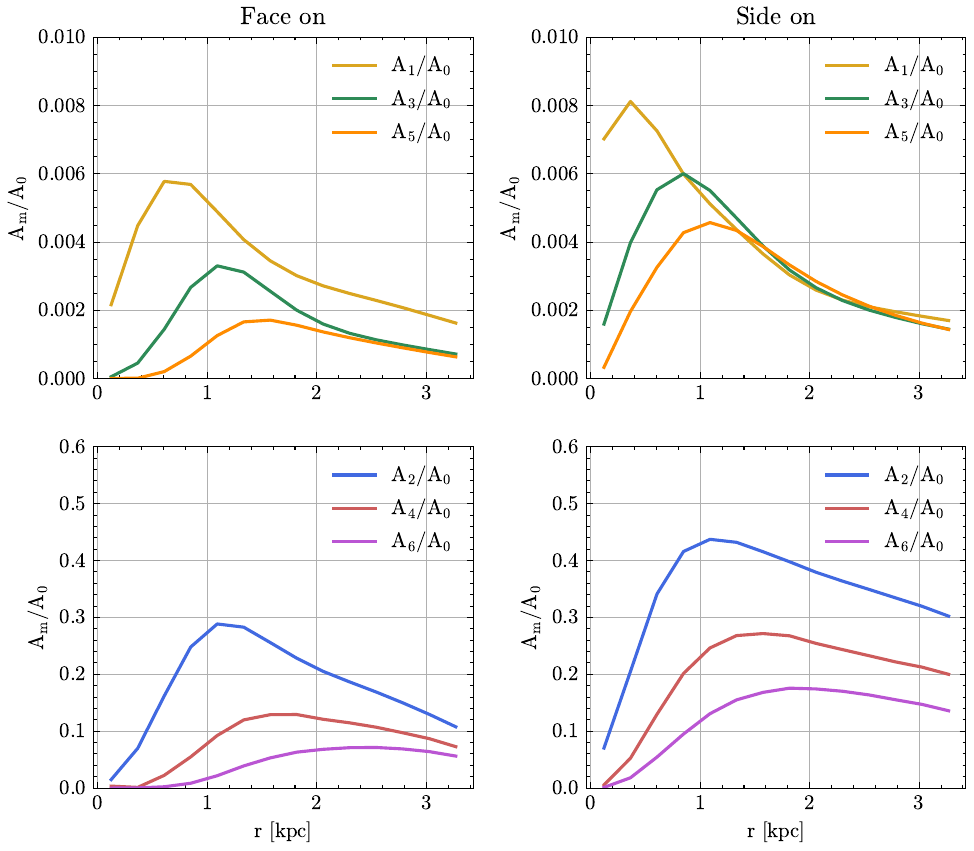} 
\caption{This figure shows the profiles of the relative Fourier amplitudes, $\mathrm{A_m/A_0}$, for Au 10 from the Auriga simulations. These profiles are measured in the bar's region, as determined in~\citet{2020MNRAS.491.1800B}. The amplitudes are shown as a function of radius in kiloparsecs (kpc). The left and right columns correspond to different galaxy projections: a face on view and an edge on view with the bar is side on, respectively. The top row displays the odd Fourier components ($m=1, 3, 5$), while the bottom row displays the even components ($m=2, 4, 6$). Within each panel, different colored lines distinguish the Fourier modes as indicated by the legend.} 
\label{fig:halo_10_A4_A2} 
\end{figure}

Figure~\ref{fig:halo_10_A4_A2} presents the radial profiles of the normalized Fourier amplitudes, $\mathrm{A_m/A_0}$, for Au 10, shown from two distinct projections: face on (left column) and side-on (right column). 
The top row displays the odd-order components ($m=1, 3, 5$), while the bottom row shows the even-order components ($m=2, 4, 6$).

Three primary trends are immediately apparent from the figure. 
First, the even-order Fourier components consistently exhibit significantly larger amplitudes than their odd-order counterparts. 
This is expected, as the galaxy's structure is dominated by the bar, a feature with strong bisymmetry ($m=2$). 
The power in higher even modes ($m=4, 6$) and all odd modes, which trace less prominent structural features or asymmetries, is naturally lower. 
Within each mode (even or odd), the amplitudes systematically decrease as the order $m$ increases. 
This reflects the physical reality that the majority of the galaxy’s structural power resides in large-scale components (like the bar itself), whereas the power in smaller-scale, higher-angular-frequency details diminishes progressively.~\citep{2002MNRAS.330...35A}.

Second, the radial location of the peak amplitude systematically shifts to larger radii as the Fourier order $m$ increases. 
This trend is a direct consequence of the spatial distribution of the physical structures that each mode traces. 
The $m=2$ mode captures the primary bar structure, with its peak amplitude occurring where the bar is most dominant—typically in the main body of the bar, inside its full length. 
In contrast, higher-order modes, such as $m=4$ and $m=6$, trace features like boxy/peanut isophotes and hexapole distortion. 
These structures are prominent toward the ends of the bar. 
Consequently, the radii where these higher-order modes have their maximum contribution are located further out than the peak of the primary $m=2$ mode.

A third notable trend is the systematic increase in the peak amplitudes of both odd and even-order modes when a galaxy is viewed edge on with side-on bars versus face on. This phenomenon is a direct consequence of projection effects tied to the bar’s intrinsic triaxial structure; as a body that is elongated within the disk plane but vertically thin, its projection appears as a more highly concentrated feature when viewed side on (see Figure~\ref{fig:halo_10} for an illustration). 
This maximizes the density contrast between the bar and the azimuthally averaged background. 
Consequently, the relative amplitudes of all non-axisymmetric Fourier modes, represented by the ratio $\mathrm{A_m/A_0}$, are enhanced.

\subsection{The correlation between the strength of the bar and lensing phenomenon}
\label{subsec:corre}

In this section, we investigate the correlation between the angular complexity in the bar region and two key metrics: the caustic area, a proxy for lensing probability, and flux-ratio anomalies.

We begin by examining Au 10 as a representative case study.  
The top panel of Figure~\ref{fig:halo_10_Cusp_caustic_relation} shows that, the tangential and radial caustics area is significantly larger for the side-on projection than for the face on view.

This trend is consistent with the structural properties of the bar itself. 
As shown in Figure~\ref{fig:halo_10_A4_A2}, the peak amplitudes of the Fourier modes are also systematically larger when the bar is viewed side-on. 
The correspondence between these two independent measurements suggests a direct correlation between the strength of the bar's non-axisymmetric features and the resulting lensing caustic area.

\begin{figure} 
\centering 
\includegraphics[width=0.6\columnwidth]{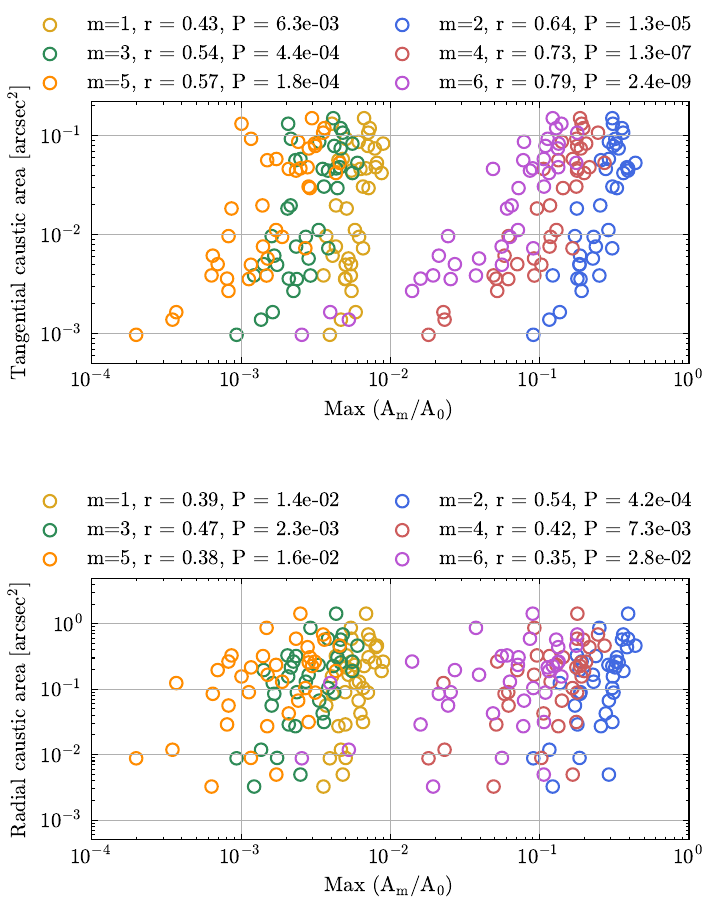}
\caption{This figure shows the correlation between the peak Fourier amplitudes in the bar region, max($\mathrm{A_m/A_0}$), representing the strength of the bar's angular complexity, and the lensing caustic area.
The top panel displays the results for the tangential caustic area. The tangential caustic has a diamond shape, as shown in Figure~\ref{fig:halo_10_Cusp_caustic_relation}, and it stretches the source image along the tangential direction.
The bottom panel displays the results for the radial caustic area. The radial caustic has an elliptical shape, as shown in Figure~\ref{fig:halo_10_Cusp_caustic_relation}, and it stretches the source image along the radial direction.
Each color represents a different Fourier component, from $m=1$ to $m=6$. 
The legend provides the Spearman correlation coefficient ($r$) and the associated p-value for each component, indicating the statistical significance of the correlation.
} 
\label{fig:boxy_strength_vs_area} 
\end{figure}

To robustly investigate these correlations, we analyze a sample of 21 barred galaxies from the Auriga simulations, identical to the sample used by~\cite{2020MNRAS.491.1800B}. 
Each galaxy is viewed from two distinct projection angles, resulting in a total of 42 test cases.
The results are presented in Figure~\ref{fig:boxy_strength_vs_area}, which plots the tangential (upper panel) and radial (lower panel) caustic areas against the peak amplitude of the Fourier modes, $\max(\mathrm{A_m/A_0})$. 
The statistical significance of these relationships is quantified using the Spearman rank correlation coefficient ($r$) and the corresponding p-value, both of which are annotated in each panel.

A clear distinction emerges between the even- and odd-order Fourier components. 
The even modes ($m=2, 4, 6$) exhibit stronger correlations with caustic area than the odd-order modes, as evidenced by their larger correlation coefficients. 
Furthermore, within both the even and odd sets, the correlation strength systematically increases with the mode number, $m$. The hexapole ($m=6$) mode, in particular, displays the strongest correlation among all components tested. 
Finally, the p-values for all six correlations are extremely low ($p < 0.01$), indicating that the observed trends are statistically significant.

These results show a direct connection between the size of the tangential caustic and the strength of higher-order angular structural components in the region of galaxy bar. 
Since these non-axisymmetric modes are all physical manifestations of the underlying bar, this finding implies a more fundamental relationship: galaxies hosting stronger, more structurally complex bars are expected to produce larger tangential caustic areas and, consequently, have a higher probability of generating strong lensing events.

In contrast, the radial caustic area exhibits a different behavior.
Here, the $m=2$ mode, representing the primary strength of the bar, shows the strongest correlation.
This suggests that the size of the radial caustic is predominantly sensitive to the bar's fundamental strength, rather than the higher-order angular structures that influence the tangential caustic.

We then investigate the correlation between the strength of the angular complexity in the bar region and the flux ratio anomaly, defined here as $\mathrm{R_{cusp}}(\Delta \phi)$ evaluated at a cusp opening angle of $\Delta \phi=60^\circ$.
This opening angle was not specially chosen; we simply wanted a straightforward way to characterize the magnitude of the lensing flux ratio anomaly. 
We also tested our results using other angles, such as 30$^\circ$ and 90$^\circ$, and found that the conclusions did not change significantly. 
It should be noted that we analyze a sample of 21 barred galaxies from the Auriga simulations, which is identical to the sample used by~\cite{2020MNRAS.491.1800B}.
We initially generated 42 lensing cases (21 galaxies $\times$ 2 projections). 
However, due to the simulation's resolution, only 21 of the 42 lensing cases resulted in image solutions that passed the magnification threshold ($\mu<100$) as introduced in Section~\ref{subsec:lens_the}.

\begin{figure} 
\centering 
\includegraphics[width=0.6\columnwidth]{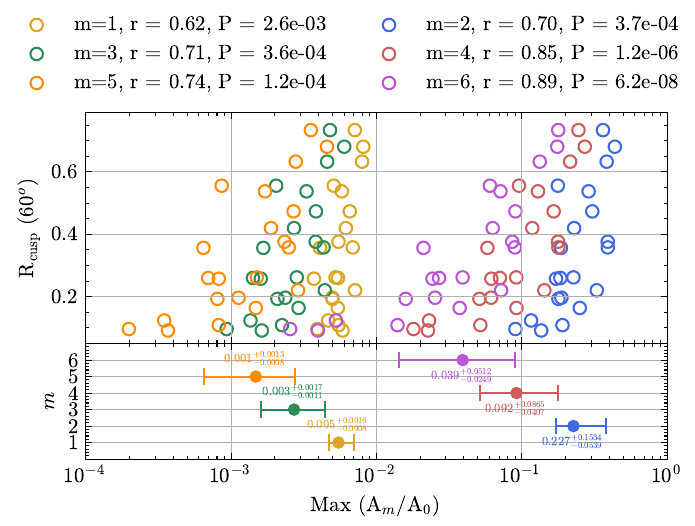}
\caption{This top panel shows the correlation between the peak Fourier amplitudes in the bar region, max($\mathrm{A_m/A_0}$), and the cusp-caustic flux ratio, $\mathrm{R_{cusp}}(60^\circ)$. The Fourier amplitudes represent the strength of the bar's angular complexity. The definition of $\mathrm{R_{cusp}}(60^\circ)$ is in Eq.~(\ref{eq:R_cusp}).
Each color represents a different Fourier component, from $m=1$ to $m=6$. 
The legend provides the Spearman correlation coefficient ($r$) and the associated p-value for each component, indicating the statistical significance of the correlation.
The lower panel shows the distribution of Fourier modes across the galaxy sample, following the same color coding. The markers and error bars represent the median value and the $1\sigma$ (16th–84th percentile) range, respectively.
}
\label{fig:boxy_strength_vs_value_at_60} 
\end{figure}

The top panel of Figure~\ref{fig:boxy_strength_vs_value_at_60} presents these relationships, with each color representing a different Fourier component, $\max(\mathrm{A_m/A_0})$, from $m=1$ to $m=6$.
The legend provides the Spearman correlation coefficient ($r$) and the associated p-value for each component, indicating the statistical significance of the correlation.
The lower panel shows the distribution of Fourier modes across the galaxy sample. The markers and error bars represent the median value and the $1\sigma$ (16th–84th percentile) range, respectively.
We find that both odd and even Fourier modes show the same trend: lower-order modes exhibit higher mean values and larger scatter.

The results in the top panel reveal several clear trends. First, the even-order components (particularly for $m=4$ and $m=6$, which capture the boxy/peanut and hexapole structures) consistently exhibit stronger correlations with $\mathrm{R_{cusp}}(60^\circ)$ than the odd-order components. Furthermore, a systematic pattern emerges within both the even and odd modes: as the Fourier order $m$ increases, the correlation with the flux ratio anomaly becomes stronger. The correlation is exceptionally strong for $m=4$ and $m=6$, with coefficients of $r = 0.85$ and $r=0.89$ and p-values on the order of $10^{-6}$ and $10^{-8}$, respectively.

This extremely tight correlation indicates a robust, physical connection between flux ratio anomalies and the presence of these higher-order features. 
Consequently, an unexpectedly high value of $\mathrm{R_{cusp}}$ may imply that the bar has boxy/peanut and hexapole components in lensed galaxies.
The presence of these structures is a key diagnostic for understanding galaxy evolution, as they are widely considered to be signposts of secular evolution processes, driven by the bar, which rearrange disk material and build up the central components of galaxies over cosmic time \citep{2004ARA&A..42..603K, 2005MNRAS.358.1477A}.

In addition to the dominant correlations found for the $m=4$ and $m=6$ modes, we also observe moderate correlations between $\mathrm{R_{cusp}}(60^\circ)$ and the peak amplitudes of the $m=1, 2, 3,$ and $5$ components. For these modes, the Spearman coefficient ($r$) can exceed $0.7$ with p-values below $3\times10^{-3}$. This suggests that an anomalously high value of $\mathrm{R_{cusp}}$ may also imply the presence of a strong bar (from the $m=2$ correlation), as well as lopsidedness ($m=1$), triangular ($m=3$), and pentagonal ($m=5$) structures within the bar of a lensing galaxy.

In conclusion, the flux ratio anomaly is strongly correlated with even, higher-order structures like the boxy/peanut ($A_4$) and hexapole ($A_6$) components, while its correlation with odd or lower-order modes is weaker.
This finding indicates that the flux ratio anomaly is particularly sensitive to small-scale, high-angular-frequency perturbations in the gravitational potential. 
It can therefore serve as a valuable and quantitative proxy for the strength of these complex, higher-order morphological features in lensed galaxies.

\section{Conclusion and discussion}
\label{sec:conclu_discus}
Bars are ubiquitous features in spiral galaxies, observed in approximately $65\%$ of the local disk galaxy population \citep{2008ApJ...675.1141S}. 
With the advent of large-scale surveys such as Euclid and the upcoming Chinese Space Station Telescope (CSST), the sample of galaxy-galaxy strong lensing systems with spiral lens galaxies is expected to grow substantially. 
Concurrently, advancements in Integral Field Unit (IFU) spectroscopic surveys and kinematic modeling provide an independent means of mapping the mass distribution of these lens galaxies. 
The combination of kinematic data with strong lensing constraints offers a powerful method for breaking the well-known degeneracy between disk and halo components~\citep{2000ApJ...533..194M, 2002MNRAS.334..621T, 2011MNRAS.417.1621D, 2012ApJ...750...10S}, leading to more accurate mass reconstructions.

In this work, we have investigated the influence of bar structures on the cusp-caustic relationship, which is a direct probe of flux ratio anomalies. 
Our analysis is based on strong lensing simulations of 21 barred spiral galaxies selected from the Auriga project, as detailed in \cite{2020MNRAS.491.1800B}. 
To isolate the impact of the bar, we developed a methodology to mitigate noise from numerical simulations (e.g., shot noise) and astrophysical substructures (e.g., dark matter subhalos). 
This was achieved by fitting the projected surface mass density, initially rendered with a Smoothed Particle Hydrodynamics (SPH) kernel, using the MGE method. 
As illustrated in Figure~\ref{fig:halo_10}, the MGE-fitted surface density is significantly smoother and more symmetric than the direct SPH projection, ensuring that our measured flux ratio anomalies are predominantly influenced by the large-scale asymmetry of the bar.

Figure~\ref{fig:halo_10_Cusp_caustic_relation} shows the lensing simulation result for the lens galaxy shown in Figure~\ref{fig:halo_10}.
The caustic area is notably larger for an edge on projection compared to a face on view, a direct consequence of the higher projected surface mass density in the edge on case. 
Furthermore, we found that the cusp relation, $\mathrm{R_{cusp}}$, is nearly linearly proportional to the opening angle of the triplet images.
We leverage this linearity to define a standardized metric for the flux ratio anomaly, $\mathrm{R_{cusp}}(60^\circ)$, derived from a linear fit to the cusp relation.

To characterize the bar's structure, we employed a Fourier decomposition of the MGE-derived surface density. 
The strengths of the various morphological components were quantified by the peak amplitude of their respective Fourier modes, $A_m/A_0$, as shown in Figure~\ref{fig:halo_10_A4_A2}. 
This analysis highlighted three key trends: (1) The galaxy's morphology is dominated by even-order modes, particularly the $m=2$ bar component. 
(2) The peak amplitudes of higher-order modes ($m>2$) occur at larger radii, tracing features such as boxy/peanut-shaped isophotes. 
(3) An edge on projection enhances the amplitudes of all modes by increasing the density contrast of the bar against the disk.

We also find statistically significant correlations between the bar's structural components and its lensing properties. 
The caustic areas, a proxy for the strong lensing cross-section, show distinct dependencies on different Fourier modes (Figure~\ref{fig:boxy_strength_vs_area}). 
The tangential caustic area correlates most strongly with higher-order even modes (specifically, the $m=6$ hexapole component), indicating that structurally complex bars produce larger tangential caustics. 
In contrast, the radial caustic area is most sensitive to the primary bar strength ($m=2$). It is well-established that the caustic area is primarily determined by the central density of the lens galaxy \citep{1992grle.book.....S}. 
Our findings therefore suggest a possible physical connection: galaxies with higher central densities tend to host stronger and structurally more complex bars.
We will investigate this connection in detail in our future work.

The central result of this paper is the strong, direct correlation between $\mathrm{R_{cusp}}$ and the strength of the bar's different components. 
We find that the boxy/peanut ($m=4$) and hexapole ($m=6$) components show exceptionally tight relationships, with Spearman's $r = 0.85$ and $0.89$, and p-values on the order of $10^{-6}$ and $10^{-8}$, respectively. 
This demonstrates that flux ratio anomalies are most significantly influenced by high-frequency angular components, which are characteristic of complex bar structures.

Furthermore, we compared the simulation predicted $\mathrm{R_{cusp}}$ distributions with the latest observational sample of quadruply lensed quasars~\citep{Xu_2015, 2024MNRAS.530.2960N, Keeley2024_JWST_II, Keeley2025_JWST_III}. 
We found that barred galaxies naturally produce systematically larger $\mathrm{R_{cusp}}$ values than those predicted for elliptical galaxies hosting dark matter subhalos. 
Crucially, the observed $\mathrm{R_{cusp}}$ values of the four known spiral lens systems fall well within the expected theoretical range (below the 50\% probability contour) from our barred galaxy simulations. 
This means that the flux-ratio anomalies in these spiral systems can be fully accounted for by the macro-structure of the galactic bar, without requiring the presence of dark matter subhalos.

Finally, one may worry that barred galaxies are often associated with dust lanes, and that dust extinction can affect the observed fluxes of lensed images. This can influence studies of galaxy structure using flux-ratio anomalies. However, most strongly lensed systems analyzed in flux-ratio anomaly studies rely on radio or mid-infrared measurements to avoid the influence of microlensing, where dust extinction is almost negligible.
In addition, our tests also show that the flux-ratio anomaly induced by the bar is almost independent of source size (as discussed in Appendix~\ref{sec:ap_finite}). As a result, the wavelength-dependent nature of flux-ratio anomalies caused by dust extinction can be distinguished from those induced by the bar.
Furthermore, this difference in the source-size dependence of flux-ratio anomalies induced by dark matter substructure~\citep{2012MNRAS.419.3414M} versus bars also provides a potential observational signature to distinguish between these two mechanisms.

In conclusion, this study, the first in a series, finds that the higher-order structural components of bars (e.g., boxy/peanut and hexapole features) are strongly correlated with flux ratio anomalies and the observed flux ratio anomaly in a system with a spiral lens galaxy may not exclusively signify the presence of dark matter subhalos, but could instead be caused by the complex structure of the bar itself. 
Consequently, our results underscore the necessity of incorporating realistic, morphologically complex bar components into strong lensing models to accurately interpret observations, especially when constraining dark matter subhalos.

\section*{Acknowledgements}
We thank the anonymous referee for constructive comments that improved this manuscript.
We have used simulations from the Auriga Project public data release~\citep{2024MNRAS.532.1814G} available at \url{https://wwwmpa.mpa-garching.mpg.de/auriga/data}. 
This work is partly supported by the National Science Foundation of China (Grant No. 12133005).
Y.J. acknowledges the support of the National Science Foundation of China (Grant No. 12403017).
X.S. acknowledges support from Shuimu Tsinghua Scholar Program (No. 2024SM199) and the China Postdoctoral Science Foundation (Certificate Number: 2025M773189).

\section*{Data Availability}
The data that support the findings of this study are available from the corresponding author upon reasonable request.



\bibliographystyle{mnras}
\bibliography{example} 




\appendix
\section{The influence of galactic evolution}
\label{sec:ap_evo}
\begin{figure} 
\centering 
\includegraphics[width=0.6\columnwidth]{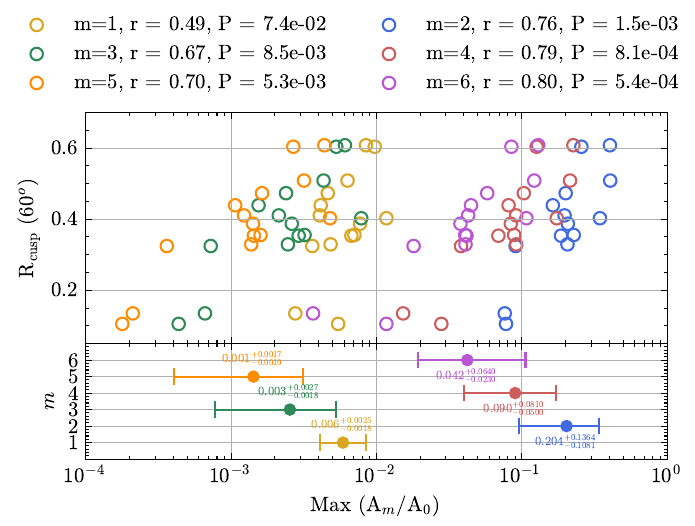}
\caption{Results obtained using the $z=0.5$ snapshot of the Auriga simulations. The top panel shows the correlation between the peak Fourier amplitudes in the bar region, max($\mathrm{A_m/A_0}$), and the cusp-caustic flux ratio, $\mathrm{R_{cusp}}(60^\circ)$. The Fourier amplitudes represent the strength of the bar's angular complexity. The definition of $\mathrm{R_{cusp}}(60^\circ)$ is in Eq.~(\ref{eq:R_cusp}).
Each color represents a different Fourier component, from $m=1$ to $m=6$. 
The legend provides the Spearman correlation coefficient ($r$) and the associated p-value for each component, indicating the statistical significance of the correlation.
The lower panel shows the distribution of Fourier modes across the galaxy sample, following the same color coding. The markers and error bars represent the median value and the $1\sigma$ (16th–84th percentile) range, respectively.
}
\label{fig:boxy_strength_vs_value_at_60_z0p5} 
\end{figure}

In the main text, we analyze samples from the $z=0$ snapshot of the Auriga simulations. In this section, we test whether our conclusions depend on galactic evolution.
Here, we run the full simulation as described in the main text, but use the snapshot at $z=0.5$.
The results are shown in Figure~\ref{fig:boxy_strength_vs_value_at_60_z0p5}.

We find that the trend reported in the main text is still observed in the $z=0.5$ configuration, where higher-order structural components of bars (e.g., boxy/peanut and hexapole features) are strongly correlated with flux ratio anomalies.
Although the p-values are larger than those reported in the main text, this is attributed to the fact that younger galaxies typically host weaker bars.
This can be seen by comparing the mean value of $\mathrm{A_2/A_0}$ in the bottom panel between Figure~\ref{fig:boxy_strength_vs_value_at_60} and Figure~\ref{fig:boxy_strength_vs_value_at_60_z0p5}, where the $z=0$ configuration shows a larger value.
As a result, the caustic area is smaller, and fewer samples pass the numerical selection described in Section~\ref{subsec:lens_the}.

\section{Finite source effects on flux ratio anomalies induced by bars}
\label{sec:ap_finite}
In this work, we assume a point-like source. However, \citet{2012MNRAS.419.3414M} found that finite source effects are important in studies of flux ratio anomalies induced by small-scale dark matter substructure, and that the commonly used point-source approximation can lead to biased results. This is because, for flux anomalies caused by dark matter subhalos, it is well established that when the source size exceeds the characteristic scale of the subhalos, the flux-ratio anomalies are effectively washed out.

\begin{figure} 
\centering 
\includegraphics[width=0.5\columnwidth]{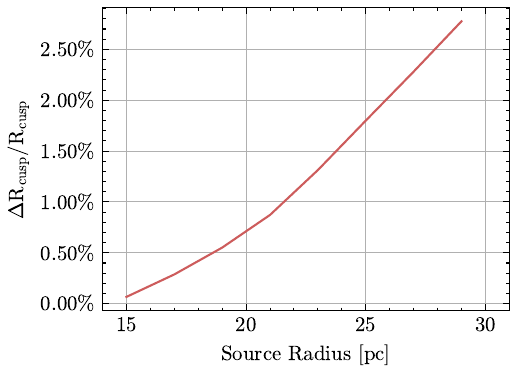} 
\caption{The x-axis represents the source size and y-axis shows the relative difference in $R_{\mathrm{cusp}}$ between the finite source and point source cases.} 
\label{fig:R_cusp_trend_127_xz} 
\end{figure}

In this section, to explicitly quantify the impact of a finite source size on our results, we compute the finite-source effect on the flux-ratio anomaly. The results are presented in Figure~\ref{fig:R_cusp_trend_127_xz}. The x-axis represents the source size; to clearly illustrate the finite-source effect, we consider source sizes ranging from $15$~pc to $\sim30$~pc, which are larger than the observed sizes of radio-emitting or narrow-line regions in quasars. The y-axis shows the relative difference in $R_{\mathrm{cusp}}$ between the finite source and the point source cases.

We find that the relative differences in $R_{\mathrm{cusp}}$ between the finite source models and the point-source limit are less than $3\%$. This demonstrates that the finite-source effect has a negligible impact on flux-ratio anomalies induced by bars and can be safely ignored for the quasars studied in this paper. Furthermore, this difference in the source-size dependence of flux ratio anomalies induced by dark matter substructure versus bars provides a potential observational signature to distinguish between anomalies caused by dark matter subhalos and those arising from bars.

\bsp	
\label{lastpage}
\end{document}